\documentclass[aps,prb,twocolumn,showpacs,preprintnumbers,amsmath,amssymb]{revtex4}
\usepackage{amssymb}
\usepackage{graphicx}
\usepackage{color}
\usepackage{dcolumn}
\usepackage{bm}

\begin{document}

\title{Phase diagram of CeFeAs$_{1-x}$P$_{x}$O obtained from
electric resistivity, magnetization, and specific heat
measurements}

\author{Yongkang Luo$^1$, Yuke Li$^1$, Shuai Jiang$^1$, Jianhui Dai$^2$, Guanghan Cao$^1$,
and Zhu-an Xu$^1$\footnote[1] {Electronic address:
zhuan@zju.edu.cn}}

\address{$^1$ State Key Laboratory of Silicon
Materials and Department of Physics, Zhejiang University, Hangzhou
310027, China}
\address{$^2$ Department of Physics, Zhejiang
University, Hangzhou 310027, China}

\date{\today}

\begin{abstract}

We performed a systematic study on the properties of
CeFeAs$_{1-x}$P$_{x}$O ($0\leq x\leq 1$) by electrical
resistivity, magnetization and specific heat measurements. The
$c$-axis lattice constant decreases significantly with increasing
P content, suggesting a remarkable chemical pressure. The Fe-$3d$
electrons show the enhanced metallic behavior upon P-doping and
undergo a magnetic quantum phase transition around $x \approx
0.4$. Meanwhile, the Ce-$4f$ electrons develop a ferromagnetic
order near the same doping level. The ferromagnetic order is
vanishingly small around $x=0.9$. The data suggest a
heavy-fermion-like behavior as $x\geq 0.95$. No superconductivity
is observed down to 2 K. Our results show the ferromagnetic
ordered state as an intermediate phase intruding between the
antiferromagnetic bad metal and the nonmagnetic heavy fermion
metal and support the cerium-containing iron pnictides as a unique
layered Kondo lattice system.

\end{abstract}

\pacs{74.62.Dh, 74.62.Bf, 74.62.Fj, 71.20.Eh, 74.70.Dd}

 \maketitle

\section{Introduction}

The homologous quaternary iron pnictides, {\it R}Fe{\it X}O, ({\it
R}: rare earths, {\it X}: As or P) show a diversity of physical
properties although they have the same ZrCuSiAs-type crystal
structure with a unique Fe{\it X}-layer sandwiched by the {\it
R}O-layer\cite{Pottgen:08Review}. Typically, LaFeAsO is an
antiferromagnetic (AFM) metal below 140 K, and is a superconductor
(SC) with maximal $T_{c}$ = 26 K upon F-doping\cite{Hosono-LaOF}.
$T_{c}$ can be surprisingly increased up to 41 K when La is
replaced by Ce \cite{WangNL-CeOF} or even higher if replaced by
other rare earths\cite{Chen-SmOF,ZhaoZX-SmOF,WangC-GdTh}.
Meanwhile, the parent compounds (denoted as Ln-1111) of these
SC's, such as CeFeAsO, are still Fe-$3d$ itinerant AFM's similar
to LaFeAsO. However, a noticeable $f$-electron AFM ordering of
Ce$^{3+}$ is also observed at a much lower temperature ($\sim$4
K)\cite{WangNL-CeOF}. On the other hand, the stoichiometric LaFePO
is a low $T_c$ SC without any trace of the AFM ordering
\cite{Hosono-LaP,DHLu-LaP}. By contrast, CeFePO is a heavy fermion
(HF) metal with Kondo temperature $T_K\sim $ 10
K\cite{Geibel-CeP}.

These discoveries put the rare-earth iron pnictides on the
boundary between the high-$T_c$
superconductors\cite{PALee-cuprates} and the heavy fermion
metals\cite{Coleman-HF}, and open an avenue for searching the
complicated interplay of various $d$- and $f$-electron
correlations. Within the iron-pnictogen layer, the $d$-electrons
are expected to be more itinerant in the phosphides than in the
arsenides\cite{Si-08PRL,Si-NJP}, with the former being recently
identified as a moderately correlated metal\cite{Qazilbash-NP}. In
contrast to the cuprates, the inter-layer distance between the
transition metals and the rare-earths in the iron pnictides is
critically significant so that the coupling between them may play
an important role even in the parent compounds of the high-$T_c$
Fe-based SC's\cite{DaiJH-CeFe}. Recent neutron scattering and muon
spin relaxation experiments indeed provide evidence for a sizable
inter-layer coupling in CeFeAsO\cite{Dai-CeCEF,Maeter-09}. So far
it is still unclear whether this coupling is due to some kind of
polarization effect raised by the ordered moment of
Fe$^{2+}$-ions\cite{Maeter-09} or more microscopically the
effective hybridization between the $3d$- and $4f$-orbitals
bridged by the pnictogens\cite{DaiJH-CeFe}. A first-principle
local-density approximation (LDA) plus dynamical mean-field theory
(DMFT) study \cite{Pourovskii-CePrNd} suggested that applying
physical pressure will enhance the $d$-$f$ hybridization, leading
to the Kondo screening of the Ce-moments. However a sufficiently
high pressure is required in order to observe this effect. It
turns out that CeFeAs$_{1-x}$P$_{x}$O, i.e., P doping at As sites
in CeFeAsO, may provide, among others, a unique layered Kondo
lattice system to probe the intriguing $3d$-$4f$ electron
interplay under ambient pressure\cite{DaiJH-CeFe}.
Chemical-pressure-induced superconductivity has been observed in
P-doped LaFeAsO \cite{WangC-LaP}.

We report a systematic study on the doping evolution of the
physical properties of CeFeAs$_{1-x}$P$_{x}$O using electrical
resistivity $\rho (T)$, magnetic susceptibility $\chi (T)$,
isothermal magnetization $M(H)$ and specific heat $C(T)$
measurements. A series of 21 different P-doped polycrystalline
samples were synthesized. Our results reveal a rich phase diagram
consisting of an AFM quantum critical point (QCP) of the Fe-$3d$
electrons and a possible ferromagnetic (FM) QCP of the Ce-$4f$
electrons. An intermediate FM phase emerges in between the AFM bad
metal and the nonmagnetic HF metal. Because P doping does not
introduce extra electrons but shortens the $c$-axis significantly,
these findings provide a rare example of chemical-pressure-induced
HF metals with strong FM fluctuations.

\section{Experimental}

We synthesized a series of CeFeAs$_{1-x}$P$_{x}$O (0$\leq x \leq$ 1)
polycrystalline samples by solid state reaction, during which, Ce,
Fe, As, P and CeO$_{2}$ of high purity($\geq$ 99.95\%) were used as
starting materials. First, CeAs (or CeP) was presynthesized by
reacting Ce discs and As (or P) powders at 1323 K for 72 h. FeAs (or
FeP) was prepared by reacting Fe and As (or P) powders at 1173 K (or
1023 K) for 20 h. Second, powders of CeAs, CeP, CeO$_{2}$, FeAs and
FeP were weighed according to the stoichiometric ratio, thoroughly
ground, and pressed into a pellet under a pressure of 600 MPa in an
Argon filled glove box. The pellet was sealed into an evacuated
quartz tube, which was then slowly heated to 1448 K and kept at that
temperature for 50 h.

Powder X-ray diffraction (XRD) was performed at room temperature
using a D/Max-rA diffractometer with Cu-K$_{\alpha}$ radiation and
a graphite monochromator. Lattice parameters were refined by a
least-squares fit using at least 30 XRD peaks, and the structural
refinements were performed using the programme RIETAN 2000. The dc
magnetization measurement was carried out in a Quantum Design
Magnetic Property Measurement System (MPMS-5). Physical Property
Measurement System(PPMS-9) was used to take the resistivity
measurement, as well as specific heat.

\section{Results and Discussion}

The room-temperature powder x-ray differaction (XRD) patterns of
CeFeAs$_{1-x}$P$_{x}$O are shown in Fig.1. All the XRD peaks can
be well indexed based on the tetragonal ZrCuSiAs-type structure
with the space group P4/$nmm$ (No.129), and no obvious impurity
phases can be detected, suggesting high quality of the samples. As
shown in Fig.1(a-b), the XRD peaks shift toward righthand,
especially for the high angle reflections. The (003) peaks shift
much faster than the (110) peaks so that they merge into one peak
in the high doping case, indicating that the $c$-axis shrinks more
severely than the $a$-axis. This observation is consistent with
the lattice parameter calculation results shown in Fig.1(c).
Fig.1(d-e) present the Rietveld refinement profiles for the two
end members CeFeAsO and CeFePO, which give details of the crystal
structure. The remarkable differences in structure lie in the
positions of Ce and As/P. For CeFePO, the thickness of FeP layers
is much smaller, meanwhile, Ce atomic layers are much closer to
the FeP layers. This structural feature supports the strong
coupling between Fe-3$d$ and Ce-4$f$ electrons.

\begin{figure*}
\includegraphics[width=17cm]{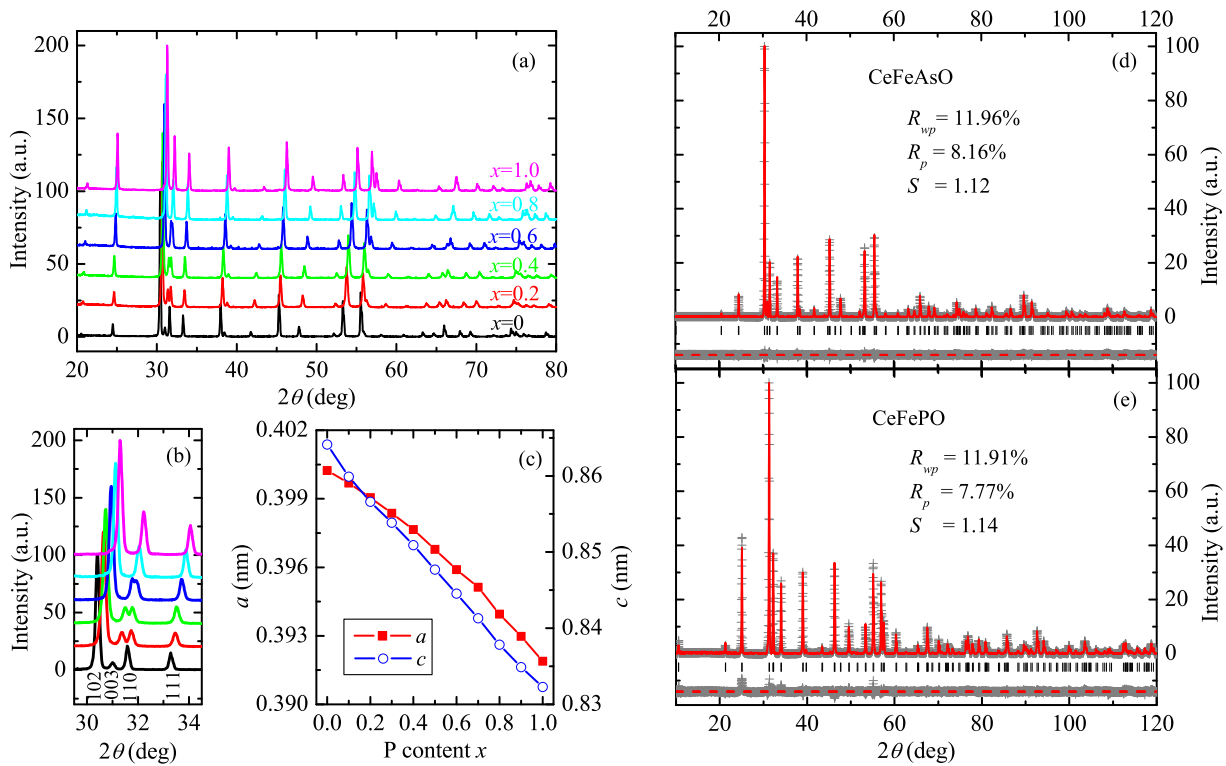}
\caption{(Color online) (a), X-ray powder diffraction (XRD)
patterns of $x$=0, 0.2, 0.4, 0.6, 0.8 and 1.0 measured at room
temperature. (b), Expanded XRD patterns showing the peak shift
with the P doping. (c), Lattice parameters as functions of P
doping level. (d-e), Rietveld refinement profiles of CeFeAsO and
CeFePO. The refined atomic coordinates ($x$, $y$, $z$) are:
Ce(0.25, 0.25, 0.1411), O(0.75, 0.25, 0), Fe(0.75, 0.25, 0.5),
As(0.25, 0.25, 0.6547) for CeFeAsO; and Ce(0.25, 0.25, 0.1508),
O(0.75, 0.25, 0), Fe(0.75, 0.25, 0.5), P(0.25, 0.25, 0.6384) for
CeFePO.}
\end{figure*}

The temperature dependent resistivity is shown in Fig.2, where four
prominent features can be identified. (i) The overall suppression of
the room temperature resistivity with increasing doping. The
resistivity at $T$ = 300 K decreases gradually from 300 $\mu\Omega$m
for $x$=0 to 12 $\mu\Omega$m for $x$=1, indicating that P-doping
enhances the metallic behavior significantly. (ii) The resistivity
anomaly, mostly pronounced for $x=0$. This anomaly was ascribed to
the structure distortion and the accompanied Fe-AFM
transition\cite{Dai-CeNeutron, Hess}. Upon doping, the anomaly is
suppressed monotonically and soon becomes an unremarkable kink at
lower temperatures. No clear kink can be identified for $x
>$ 0.3. The suppression of this anomaly with P-doping is consistent with the suppression of AFM order of Fe ions, which is confirmed by a recent neutron study\cite{DaiPC-CeP}. We mark arrows in Fig.2(a) to show the Fe-AFM
transition ($T_{N1}$, data from Ref.\cite{DaiPC-CeP}). (iii) The
resistivity upturn at low temperatures for $x\leq 0.4$. The upturn
behavior is suppressed by doping or magnetic field, see the inset
in Fig.2(b). (iv) The metallic behavior in the entire measured
temperature range for $x>0.4$. In particular, the resistivity
drops rapidly at low temperatures for larger $x$. Such a drop in
resistivity has already been observed in the HF metal CeFePO in
the previous report\cite{Geibel-CeP}. However, no SC is observed
down to $T$=2 K for the entire P doping range.

It should be noted that the value of resistivity itself may not
reflect the poly-crystalline sample's quality. Actually the most
likely impurity phases in the Ln-1111 pnictides could be FeAs or
Fe$_2$As, and Ln$_2$O$_3$. Both FeAs and Fe$_2$As are highly
metallic, more metallic than CeFeAsO itself at low temperatures.
The metallic impurity phase like FeAs could distribute most likely
in the grain boundaries and thus result in a lower resistivity if
the main phase is a bad metal. The intrinsic resistivity behavior
of Ln-1111 parent pnictides below the Fe-AFM order is still
ambiguous. For example, the resistivity of single crystal of
LaFeAsO shows an upturn at low temperatures\cite{La-R}, meanwhile
the resistivity of both poly-crystalline and single-crystalline
CeFeAsO samples shows clear metallic behavior\cite{WangNL-CeOF,
Ce-R}. However, our more than 20 CeFeAs$_{1-x}$P$_x$O samples with
different P content were prepared under same conditions and the
XRD patterns show that they all are of same high purity, the
systematic decrease in resistivity indeed suggests the increasing
metallicity with increasing P content.

\begin{figure}
\includegraphics[width=8cm]{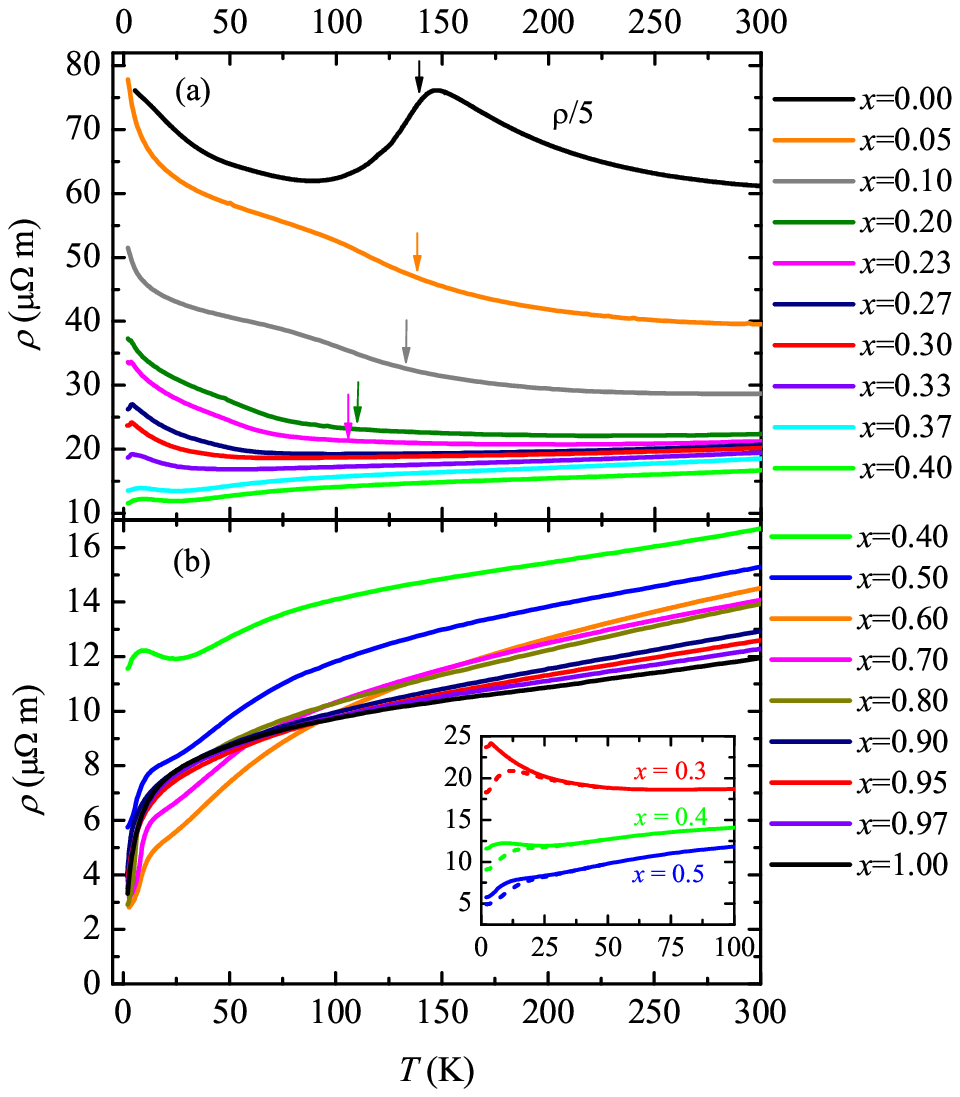}
\caption{(Color online) Temperature dependence of resistivity in
CeFeAs$_{1-x}$P$_{x}$O. (a), For $0 \leq x \leq 0.4$. Arrows
denotes the temperature of Fe-AFM order, $T_{N1}$, which were
taken from Ref.\cite{DaiPC-CeP}. (b), For $0.4 \leq x \leq 1$. The
inset shows the resistivity of $x$=0.3, 0.4 and 0.5, under zero
field (solid lines) and $\mu_0H$ of 5 T (dashed lines).}
\end{figure}

The temperature dependence of dc magnetic susceptibility is shown
in Fig.3. In the high temperature range, $\chi (T)$ increases with
decreasing temperature down to 150 K following a Curie-Weiss law
for all $0\leq x\leq 1$. The derived effective moment $\mu_{eff}$
is 2.54 $\mu_{B}$ for $x$ = 0, and 2.56 $\mu_{B}$ for $x=1$
[insets of Fig.3], very close to that of a free Ce$^{3+}$ ion,
2.54 $\mu_{B}$. By lowering temperatures, a clear peak was
observed at 4.16 K for $x=0$, related to the formation of the
Ce$^{3+}$ AFM order. With increasing $x$, $\chi(T)$ increases and
the peak becomes round (or shoulder-like); meanwhile the
corresponding AFM ordering temperature does not change too much.
For $x\gtrsim 0.4$, an obvious divergence between
zero-field-cooling (ZFC) and field-cooling (FC) is seen at lower
temperatures [see Fig.3(b)], manifesting the FM ordered ground
state of the Ce-$4f$ electrons. The suppression of AFM and
development of FM are consistent with the enhanced metallic
transport behavior in the course of P-doping. The FM ordered state
persists until $x \sim$ 0.9 where the divergency of $\chi(T)$
vanishes.

\begin{figure}
\includegraphics[width=8cm]{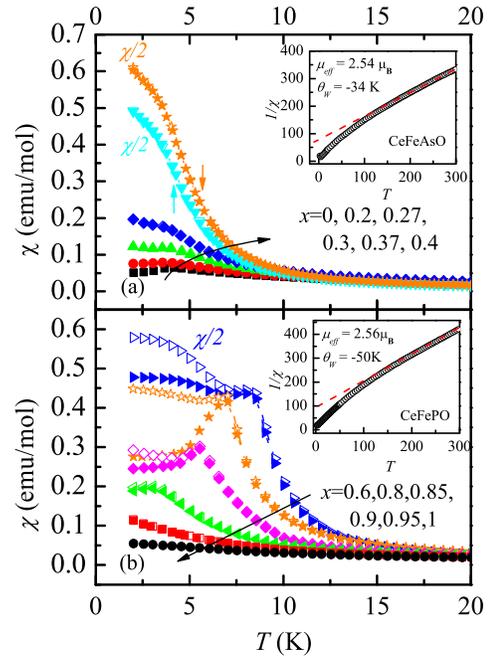}
\caption{(Color online) Temperature dependence of dc magnetic
susceptibility under ZFC (solid symbols) and FC (open symbols)
protocols for CeFeAs$_{1-x}$P$_{x}$O samples under a magnetic field
of 1000 Oe. Arrows are guides to eyes to indicate the FM transition
temperature $T_{C}$ which is defined from the extrapolation of
$1/\chi-T$ plot. The insets show the Curie-Weiss behavior of
susceptibility above 150 K for CeFeAsO and CeFePO respectively.}
\end{figure}

The evolution of Ce$^{3+}$ magnetism can be further demonstrated
by isothermal field-dependent magnetization measurements, which
are shown in Fig.4. The temperatures are fixed at 2 K and 10 K,
respectively. For $x=0$ and $T$=2 K, a kink in the magnetization
curve is clearly seen, indicating a possible spin flop in the AFM
state of Ce$^{3+}$. When $x$ increases the kink becomes weaker. As
$x \geq 0.4$, $M(H)$ starts to increase rapidly for small fields
and tends to saturate at higher fields, i.e., it exhibits a
nonlinear field dependence, see Fig.4(d) for $x=0.4$. For
$0.5\lesssim x <$ 0.9, a clear hysteresis loop can be observed and
the saturated $M$ value reaches the maximum around $x = 0.6$ [see
Fig.4(e)]. Therefore, the long-range FM order exists in a wide
doping range in between $x\sim 0.4$ and $x\sim 0.9$. The largest
saturated magnetic moment is about 0.95 $\mu_{B}$, close to 1
$\mu_B$ expected for a Ce$^{3+}$ doublet ground
state\cite{Geibel-CeP,Dai-CeCEF}. For $0.9\lesssim x< 0.95$ the
hysteresis loop is hardly observable down to the lowest measured
temperature while the saturation tendency persists. This may be
due to strong FM fluctuations at the quantum critical point
associated with the $4f$-electrons. On the other hand, both the
susceptibility and magnetization show a non-magnetic HF-like phase
for $x \geq$ 0.95 [see Fig.3(b) and Fig.4(g-h)], consistent with
the resistivity measurement.

\begin{figure*}
\includegraphics[width=1.9\columnwidth]{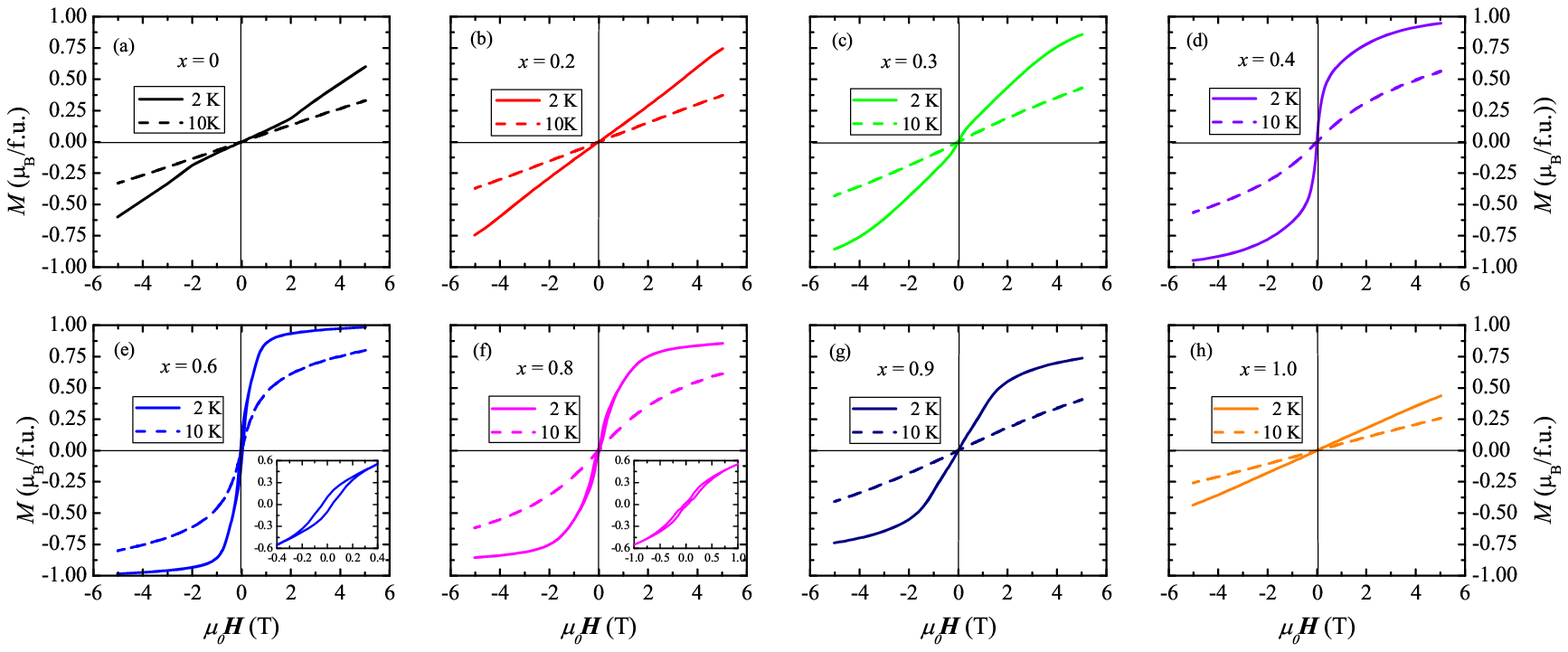}
\caption{(Color online) Isothermal magnetization of
CeFeAs$_{1-x}$P$_{x}$O. (a), $x$ = 0. (b), $x$ = 0.2. (c), $x$ =
0.3. (d), $x$ = 0.4. (e), $x$ = 0.6. (f), $x$ = 0.8. (g), $x$ = 0.9.
(h), $x$ = 1.0. For clarity, only the data of 2 K (solid lines) and
10 K (dashed lines) are shown. The insets of (e) and (f) display the
magnetic hysteresis loop at 2 K.}
\end{figure*}

The metallic HF-like behavior can be further manifested by the
specific heat measurement as shown in Fig.5. For $x=0.95, 0.97$,
and $1$, the specific heat coefficient $\gamma(T)=C(T)/T$ shows
$\log T$ behavior at low temperatures down to 2 K, see the inset
of Fig.5. We define a characteristic temperature $T_{\gamma}$
(below which HF-like behavior appears) as the intersection of two
extrapolate lines of the specific heat data. For the $x=1$ sample,
$T_{\gamma}$ $\sim$ 8.6 K. The saturated $\gamma$ should be about
700-800 mJ/molK$^{2}$ or larger, in agreement with the result of
Ref.\cite{Geibel-CeP} where $\gamma$ increases in proportion to
$\log T$ before it gets saturated below 1 K. As a comparison,
$\gamma(T)$ shows a $\lambda$-peak maximized around 4.2 K and 6.8
K in the AFM and FM ordered phases, for $x$ =0 and 0.8,
respectively, in agreement with the magnetization measurement. A
sudden increase in the saturated $\gamma$ close to $x\sim 0.9$ can
be inferred by the values of $\gamma(T = 2$K), accompanied by a
decrease in the saturated magnetization under $\mu_0 H$ = 5 T.
This implies the enhancement of the Kondo screening near the FM
instability. It should be noted that a very tiny kink was observed
around 6 K on the $C/T-T$ curve of $x$=0.95. We ascribe this
additional peak to the contribution from some impurities with
lower P content. Roughly estimated from the intensity of the kink,
the amount of the impurity phase should be less than 2\%.

We summarize the experimental results by suggesting an electronic
phase diagram shown in Fig.6. At low temperatures, both the
Fe-$3d$ and Ce-$4f$ electrons show the long-range AFM ordering for
$x\lesssim 0.37$. The $d$-electrons are Pauli paramagnetic(PM) for
$x\gtrsim0.4$, while the $f$-electrons are FM ordered until $x\sim
0.9$. For $x\geq 0.95$, the $f$-moments are completely quenched
and the whole system becomes a nonmagnetic HF-like metal. The
HF-like region is roughly depicted by $T_{\gamma}$. Therefore, in
addition to the first magnetic QCP (denoted by $x_{c_{1}}$)
associated with the AFM-PM transitions of the $d$- electrons,
there could be a another QCP associated with the FM-HF transiton
of $f$-electrons around $x$ of 0.92 (denoted by $x_{c_{2}}$).
$x_{c_{1}}$ has been known to be ~0.4 by the study of neutron
scattering\cite{DaiPC-CeP}. A turning point (denoted by
$x_{c_{3}}$) associated with the AFM-FM transition of the
$f$-electrons at $x_{c_3}\sim 0.37$ is quite close to $x_{c_1}$.
Meanwhile the strong FM fluctuations show up in the vicinity of
$x_{c_2}$.

It should be noticed that the transition temperatures of the
Fe-AFM ordering and the structure distortion can be approximately
determined from the resistivity anomaly as shown in Fig.2 and
neutron studies\cite{DaiPC-CeP}. They are undistinguishable in
resistivity when $x>0.2$. The neutron study\cite{DaiPC-CeP}
revealed that both of them are suppressed at $x_{c_{1}}\sim 0.4$,
confirming an early theoretical prediction for the existence of a
unique magnetic QCP driven by P-doping\cite{Dai-PNAS}. Moreover,
the $d$-electron QCP seems to be very close to the turning point
($x_{c_3}$) of the $f$-electrons. This fact may indicate that the
$d$-electron magnetic transition is of weak first order
accompanied by a reconstruction of the electronic states across
$x_{c_1}$. It is not certain if the coincidence implies the AFM
ordering in Ce subsystem being supported by the AFM order in the
Fe plane, because the ordering patterns of the two subsystems are
different and the $f$-electron AFM order can coexist with the
$d$-electron SC order in the F-doping
case\cite{WangNL-CeOF,Dai-CeNeutron}. Besides this puzzling
feature, other implications of our experimental results should be
addressed as follows.

\begin{figure}
\includegraphics[width=8cm]{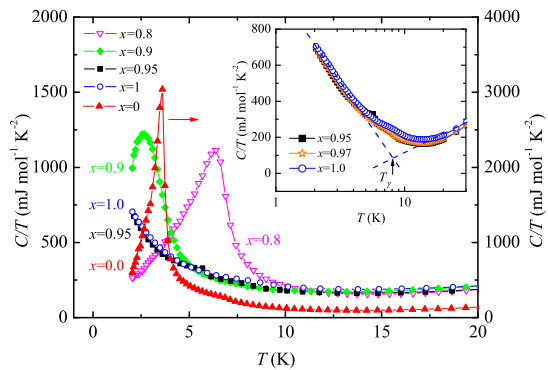}
\caption{(Color online) Specific heat measurements of
CeFeAs$_{1-x}$P$_{x}$O. A $\lambda$-peak can been observed around
$T$ = 4.2 K and 6.8 K for $x$ = 0 and 0.8, respectively,
corresponding to the AFM and FM transitions.  Inset: For $x$ =
0.95, 0.97 and 1, the specific heat coefficient $\gamma(T)=C(T)/T$
shows $\log T$ behavior at low temperatures down to 2 K, and tends
to saturate to $\gamma (0)$ when $T\rightarrow 0$. The $\gamma
(0)$ value was extrapolated to be about 700-800 mJ/K$^{2}$mol or
larger for $x$ = 1.0.}
\end{figure}

First, as the suppression of the $a$-axis is only a fraction of
that of the $c$-axis upon P-doping, the electronic states within
the Ce-O layer should not change significantly. This is also
supported by the non-magnetic LDA calculation from which the
different Fermi surfaces of CeFeAsO and CeFePO can be mainly
ascribed to the pnictogen height
\cite{Pourovskii-CePrNd,Vildosola-P}. Therefore the HF behavior in
the end compound, CeFePO, is originated unambiguously from the
inter-layer coupling between the Fe-$3d$ and Ce-$4f$ electrons.
Second, the $f$-electron AFM-FM transition could be explained by
assuming the existence of a bare $d$-$f$ Kondo coupling ($J_K$) in
the ordered phases, as it can mediate the indirect
Ruderman-Kittel-Kasuya-Yosida (RKKY) interaction ($J_{RKKY}$)
between the Ce-$4f$ local moments. Intuitively, the exchange
interaction between the local $f$-spins , $J_{f}$, is given by
$J_{f}=J^{(0)}_{f}+J_{RKKY}$, with $J^{(0)}_{f}$ being the
exchange interaction in the absence of $J_K$, and $|J_{RKKY}|\sim
J^2_K N_{F} $, with $N_F$ being the density of states at Fermi
level (of mainly $d$-electron characteristic). $J_{RKKY}$ is
negative if $N_{F}$ is small, as it should be since the parent
compound CeFeAsO is a bad metal\cite{Si-08PRL,Si-NJP}. Upon P
doping, both $J_K$ and $N_F$ should increase, leading to a sign
change in $J_f$. Third, the negative $J_{f}$ is also the driving
force for the strong FM fluctuations in the vicinity of the
critical point $x_{c_2}$, where the Kondo scale $T_K$ (which
increases exponentially with $J_K N_F$) starts to dominate over
the magnetic ordering energy scale, $T_f\approx |J_f|$.

\begin{figure}
\includegraphics[width=9cm]{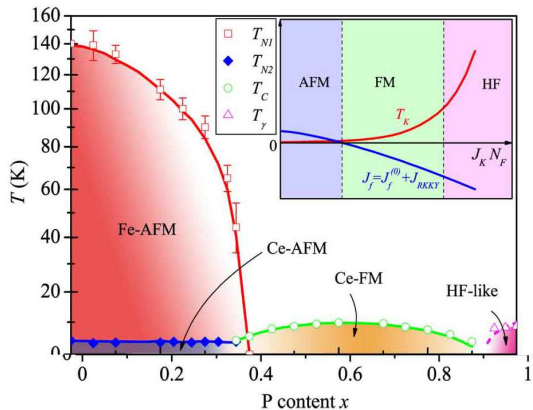}
\caption{(Color online) Electronic phase diagram of
CeFeAs$_{1-x}$P$_{x}$O (0 $\leq x \leq$ 1)). The shaded region (in
red) in the left side shows the antiferromagnetic ordering of
Fe-3$d$ moments (denoted by Fe-AFM below
$T_{N1}$\cite{DaiPC-CeP}). In the lower temperature portion of
this region, simultaneously, Ce-4$f$ moments order
antiferromagnetically (marked with Ce-AFM) below $T_{N2}$. The
small pink area at the lower right corner displays the nonmagnetic
state with heavy fermion (HF) behaviour. In the lower middle
regime (in green), the 4$f$-moments are ferromagnetically ordered
(labeled by Ce-FM) below $T_{C}$. Inset: The schematic competition
between $J_f$ and $T_K$ accounting for the the AFM-FM-HF evolution
of the $f$-electrons.}
\end{figure}

It is well-established that the RKKY interaction between the local
moments is mediated by the Kondo-coupling between the itinerant
charge carries and local moments \cite{Doniach,Hewson}. The Kondo
coupling can lead to both the Kondo effect and the RKKY
interaction, with the latter being responsible to the magnetic
ordering of the local moments. The crucial point is that the
dependences on the Kondo coupling (together with the DOS of the
charge carries) of the Kondo energy scale (describing the Kondo
effect) and of the RKKY interaction (characterizing the ordering
temperature of the local moments) are different, as described by
the intuitive formulas in the inset of Fig. 6. Based on the
similar intuitive formulas, Doniach first suggested that a quantum
phase transition from magnetic ordering phase to the paramagnetic
HF phase could exist in a Kondo lattice\cite{Doniach}. Here, we
suggest that the AFM-FM-HF evolution of the $f$-electrons in
CeFeAs$_{1-x}$P$_x$O is due to the competition between $T_K$ and
$J_f$ (instead of $J_{RKKY}$ in the conventional Kondo
lattice\cite{Coleman-HF}), see the inset in Fig.6. Notice that
$J^{(0)}_{f}$ is positive and relatively small in the absence of
$J_{RKKY}$. Thus in the AFM phase $T_K$ is mainly suppressed by
the AFM ordering gap of the $d$-electrons \cite{DaiJH-CeFe}. This
picture provides a natural explanation for the strong FM
fluctuations in a class of homologous HF compounds including
CeFePO\cite{Geibel-CeP} and CeRuPO\cite{Krellner-07,Krellner-08}.
Therefore the observed f-electron phase diagram of
CeFeAs$_{1-x}$P$_x$O system is consistent with the Doniach
picture\cite{Doniach}. In addition, it also predicts a possible
weak Kondo phase in a narrow doping region enclosing the turning
point $x_{c_3}$ (at very low temperatures). As $x_{c_3}$ is very
close to $x_{c_1}$, it may account for the observed suppression of
the SC near the AFM QCP. The existence of this peculiar phase and
its possible connection with the transport and magnetic properties
still need to be clarified.

\section{Conclusion}

In summary, our experimental study on CeFeAs$_{1-x}$P$_{x}$O ($0\leq
x\leq 1$) reveals a rich phase diagram which consists of an AFM
quantum phase transition of the Fe-$3d$ electrons and an AFM-FM-HF
evolution of the Ce-$4f$ electrons. The AFM QCP of the $d$-electrons
is very close to the turning point of the AFM-FM transition of the
$f$-electrons, while the $d$-$f$ hybridized HF state is separated by
an $f$-electron FM instability. In contrast to the F-doping case
where the high-$T_c$ SC is induced, no SC is observed down to 2 K in
the present case, signalling the unique role of P-doping in
suppressing the $d$-electron correlation\cite{Dai-PNAS,Si-NJP}.
These results highlight the importance of the inter-layer Kondo
physics\cite{DaiJH-CeFe,Coleman-09PRL} and the interplay of the $d$-
and $f$-electron correlations in the rare-earth iron pnictides.

\section*{Acknowledgments}

We thank Xi Dai and Huiqiu Yuan for helpful discussions. This work
is supported by the National Science Foundation of China, the
PCSIRT (IRT-0754), and the National Basic Research Program of
China (Grant Nos 2007CB925001 and 2009CB929104).

\end{document}